\documentclass[cits]{JINST}

\usepackage{amsmath}
\usepackage{verbatim}

\title{Symplectic integrator for $s$-dependent static magnetic fields based on
mixed-variable generating functions}

\author{A. Wolski$^{a,c}$\thanks{Corresponding author.},
J. Gratus$^{b,c}$ and
R. Tucker$^{b,c}$\\
\llap{$^a$}Department of Physics, University of Liverpool, \\ Liverpool, UK\\
\llap{$^b$}Physics Department, Lancaster University, \\ Bailrigg, Lancaster, UK\\
\llap{$^c$}The Cockcroft Institute, \\ Daresbury, Warrington, UK\\
  E-mail: \email{a.wolski@liverpool.ac.uk}}

\abstract{We describe a technique for constructing a symplectic transfer map for a charged particle
moving through an accelerator component with arbitrary three-dimensional static
magnetic field.  The transfer map is constructed by symplectic integration; by representing
the map at each step of the integration by a mixed-variable generating function,
exact symplecticity is ensured.  By using an appropriate integration algorithm, there
is no necessity to make the paraxial approximation.  The technique is illustrated by application
(in one degree of freedom) to a quadrupole magnet with strong octupole component
and fringe field.}


\preprint{Cockcroft-2012-04}

\begin{document}

%\includegraphics{CI_logo_small.eps}
%--------------------------------------------------------------------------------------------------------

\section{\label{sec:introduction}Introduction}

The ability to compute accurately and efficiently the trajectory of a particle in an
accelerator beam line is fundamental to accelerator design and operation.  Where
the beam line includes components with complex electromagnetic fields, and where
fast tracking and a high degree of accuracy are required, this problem presents
significant challenges.  Considerable work over the years has led to the development
of a variety of techniques for particle tracking (see, for example,
\cite{brown, forest1, berz2, forest2, dragt1}); however, since the equations
describing particle trajectories cannot be solved exactly, various approximations
are always needed.  Often, there is a compromise that must be made
between accuracy and speed.

There are two approaches to tracking a particle through the
electromagnetic field in a given accelerator component: either, the
equations of motion may be integrated numerically, or, a transfer map
may be applied for the entire component.  Although the equations of
motion are well known, numerical integration is in general too slow
for long-term particle tracking in a storage ring.  A transfer map,
consisting of a set of equations relating the particle position and
momentum at the exit of the component to the position and momentum at
the entrance, provides a much more efficient tracking method; however,
the problem then is to construct with good accuracy the transfer map for a
given component.  One approach is to integrate the equations of motion
using a differential algebra (DA) code \cite{differentialalgebra}, to produce
the transfer map in the form of a multi-variable power series.
Although a map in this form is very convenient for particle tracking,
in general the power series must be truncated at some order in each
step of the integration, to keep the number of terms manageable.  As a
consequence of the truncation, the final map will not be symplectic:
this may have an impact on detailed aspects of the particle dynamics
(for example, whether or not a particle remains in a storage ring over
a large number of turns \cite{yan1, kleiss1, berg2}).  Techniques have been
devised to construct symplectic, finite-order Taylor maps (Cremona maps)
that correspond to a symplectic Taylor map that has been truncated at some
order \cite{abell1, abell2}.  However, these methods are not especially
easy to implement, and the consequences of modifying a map to make it
symplectic are not always clear \cite{forest2}.

The need for efficient symplectic tracking tools has motivated the development
of techniques for manipulating symplectic transfer maps (symplectomorphisms)
using, for example, representations commonly known as Lie transformations
\cite{dragt1, lietransformations}.  Venturini and Dragt have shown how to
construct the transfer map for an accelerator component in the form of a Lie
transformation, using detailed magnetic field data \cite{dragtgeneralisedgradients}.
While Lie transformations provide a
valuable analytical framework, it is not straightforward to implement
the relevant formulae in an accelerator simulation code.  As an
alternative, we discuss here the construction of a mixed-variable
generating function \cite{goldstein}, in particular, to represent the dynamics
in an accelerator component with some specified static inhomogeneous
magnetic field.  Mixed-variable generating functions are already used
for particle tracking where symplecticity is important \cite{yan1, berg2,
berz1, yan2, berg1, warnock1, erdelyi1, erdelyi2, erdelyi3}.
Besides ensuring symplecticity, mixed-variable generating functions also
have potential advantages over Taylor maps, in reducing the number of
coefficients needed to describe the map, and in improving the accuracy
of a map for a given accelerator configuration that is obtained by
interpolation between known maps representing a set of ``reference''
configurations \cite{giboudot1,giboudot2}.
However, the common approach to using mixed-variable generating
functions  is first to construct truncated multi-variable
Taylor expansions for the canonical variables (approximating the transfer map,
usually for an entire turn of a storage ring) and then to construct
the generating function that gives this map to some desired order.
Such a process can be used to ``symplectify'' a truncated power
series; but then the same concerns apply as in the case of construction
of Cremona maps from truncated power series.
A more direct approach, that we consider here, consists of
working entirely ab-initio with mixed-variable generating functions. The
advantage is an improvement in efficiency and accuracy that can be
achieved by avoiding conversions between different (and, in some
cases, non-symplectic) representations of the same map.

Although the manipulation of maps in the form of mixed-variable
generating functions is not trivial, it is at least
possible to obtain from the generating function an exact set of
algebraic equations relating the initial and final values of the
dynamical variables that preserves the symplectic structure. 
Unfortunately, these equations then need to be
solved numerically for given initial conditions, but this can be done
to any desired precision with reasonable efficiency, using an
iterative technique such as Newton's method.  The symplecticity of the
map can then be preserved, to machine precision.

The above arguments suggest that a technique for constructing the
generating function for a given accelerator component with specified static
magnetic field offers a potentially useful tool in particle
tracking: we describe such a technique in Section\,\ref{symplecticintegrator}.
We see that relatively straightforward expressions can be written down for
the mixed-variable generating function for each integration step in the case
that the paraxial approximation can be made for the Hamiltonian
(Section \ref{sec:wfrintegrator}); slightly more complicated (though still
very manageable) expressions can be used where a more accurate form of 
the Hamiltonian is needed (Section \ref{sec:rungekutta}).
A key step in the integration is the composition of
two generating functions: this is discussed in some detail in Sections
\ref{mvgfcomposition1} and \ref{mvgfcomposition2}, where we present
two methods for performing the composition.
Then, in Section \ref{example}, we illustrate the techniques developed
in Section \ref{symplecticintegrator} by applying them to the case
of a magnetic quadrupole with strong octupole component and fringe field.

%--------------------------------------------------------------------------------------------------------

\section{Symplectic integrators\label{symplecticintegrator}}

We shall consider the case of a charged particle moving through a
static magnetic field $\vec{B}=\nabla\times\vec{A}$. Let $s$
parameterise the reference trajectory, which by assumption is an orbit
in the magnetic field. The canonical variables on the (extended) phase
space are $(\vec{x},s,\vec{p},p_s)$ where $\vec{x} = (x, y, z)$ and
$\vec{p} = (p_x, p_y, p_z)$.
$x$ and $y$ denote the position of the particle in a plane perpendicular
to the reference trajectory at $s$, which the particle crosses at time $t$;
and $z = s/\beta_0 - ct$ where $\beta_0$ is the speed (as a fraction of the
speed of light) of a particle with a fixed reference (kinetic) momentum $P_0$.
The transverse components of the canonical momentum are defined by:
\begin{subequations}
\begin{eqnarray}
p_x & = & \frac{\gamma\big(|\vec{\beta}|\big) m \dot{x} + q A_x(\vec{x},s)}{P_0}, \\
p_y & = & \frac{\gamma\big(|\vec{\beta}|\big) m \dot{y} + q A_y(\vec{x},s)}{P_0},
\end{eqnarray}
\end{subequations}
where $\gamma\big(|{\vec{\beta}}|\big)$ is the relativistic factor; the particle
velocity is $\vec{\beta}c = (\dot{x}, \dot{y}, \dot{s})$; $m$ and $q$ are
the mass and charge of the particle; and the dot indicates a derivative with respect to time.
The momentum conjugate to the longitudinal co-ordinate $z$ is the energy
deviation $p_z$, defined by:
\begin{equation}
p_z = \frac{E}{P_0 c} - \frac{1}{\beta_0},
\end{equation}
where $E = \gamma\big(|\vec{\beta}|\big) m c^2$ is the kinetic energy of the particle.
The reference momentum can
be written as $P_0 = \beta_0 \gamma_0 m c$, where $\gamma_0 = \gamma (|\beta_0|)$.
Note that third component of the particle
co-ordinate vector $\vec{x} = (x, y, z)$ actually specifies the time at which a particle
arrives at a given longitudinal position within the magnetic field; so for a magnetostatic
field, $\vec{A}$ is independent of $z$.

It is convenient
to work with the normalised vector potential, defined by:
\begin{equation}
\vec{a} = \frac{\vec{A}}{B\rho},
\end{equation}
where $B\rho = P_0/q$ is the beam rigidity and the vector $\vec{a}$
has Frenet components $(a_x, a_y, a_s)$.

In these variables, the Hamiltonian for particle motion in a static magnetic field
with a straight reference trajectory is:
\begin{equation}
H = \frac{p_z}{\beta_0} - \sqrt{\left( \frac{1}{\beta_0} + p_z \right)^2 - (p_x - a_x)^2 - (p_y - a_y)^2 - \frac{1}{\beta_0^2 \gamma_0^2}} - a_s + p_s,
\end{equation}
where $\gamma_0 = \gamma(\beta_0)$.  The independent variable is
denoted by $\sigma$.  Note that:
\begin{equation}
\frac{ds}{d\sigma} = \frac{\partial H}{\partial p_s} = 1,
\end{equation}
and therefore:
\begin{equation}
s  = \sigma + \sigma_0,
\end{equation}
where $\sigma_0$ is an (arbitrary) constant of integration.

The method will be illustrated for a particle
that moves with $y = p_y = p_z = 0$, i.e. we consider only the evolution of
the variables $x, p_x$ and $s$. The generalisation to the full number of degrees
of freedom is straightforward, and requires no
significant new features.  In addition, we assume that the particle is ultra-relativistic,
so that we can take the limit $\gamma_0 \to \infty$.
By making an appropriate gauge transformation,
it is always possible to work in a gauge in which one component of the vector
potential vanishes at all positions in space: we shall choose $a_y = 0$.
The Hamiltonian is then:
\begin{equation}
H = - \sqrt{1 - (p_x - a_x)^2} - a_s + p_s. \label{hamiltonian1dof}
\end{equation}
Our assumption that $y = p_y = 0$ as the particle moves through the field requires that:
\begin{equation}
\left. \frac{\partial a_x}{\partial y} \right|_{y = 0} = 
\left. \frac{\partial a_s}{\partial y} \right|_{y = 0} = 0.
\end{equation}
This is a restriction on the field, but again the generalisation to include other
cases requires no significant new features.

An accelerator component is described by specifying the components of the vector
potential.  Neglecting radiation and collective effects,
the transfer map for a particle in a magnetostatic field is symplectic.  Therefore,
the values of the canonical variables at the exit of the accelerator component can be
related, at least in principle, to the values of the variables at the entrance by a
mixed-variable generating
function.  Following Goldstein's nomenclature \cite{goldstein}, with
$(x_1,p_{x1})$ the entrance variables and $(x_2,p_{x2})$ the exit
variables, we write $F(x_1,p_{x2})$ as a generating
function of the second kind.  Then, the entrance and exit variables are related by:
\begin{subequations}
\begin{eqnarray}
x_2 & = & \frac{\partial F}{\partial p_{x2}} \\
p_{x1} & = & \frac{\partial F}{\partial x_1},
\end{eqnarray}
\end{subequations}
Our objective is to develop a procedure for
constructing $F(x_1,p_{x2})$ in a given static magnetic field.

%--------------------------------------------------------------------------------------------------------

\subsection{Integrator for the Hamiltonian in the paraxial approximation\label{sec:wfrintegrator}}

Assuming that the dynamical variables and normalised vector potential have small
values, we can make the paraxial approximation for the Hamiltonian given by Eq.\,(\ref{hamiltonian1dof}):
\begin{equation}
H \approx -1 + \frac{1}{2} (p_x - a_x)^2 - a_s + p_s.
\label{hamiltonianparaxial1dof}
\end{equation}
In this case, a method for constructing a convenient symplectic integrator has been
described by Wu, Forest and Robin \cite{wuforestrobin}.
A more general symplectic integrator, not requiring the paraxial approximation,
is described in Section \ref{sec:rungekutta}. However, where it is possible to make the
paraxial approximation without losing the desired level of accuracy, the advantage of doing
so is that integration can be performed more simply.
Following \cite{wuforestrobin}, using the Hamiltonian in the paraxial approximation
Eq.\,(\ref{hamiltonianparaxial1dof}), the integration of the particle motion through
a step $\Delta\sigma$ (corresponding to a step $\Delta s = \Delta\sigma$ in the field)
can be expressed as a sequence of transformations, for example:
\begin{equation}
\mathcal{M} = \mathcal{M}_1 \circ \mathcal{M}_2 \circ \mathcal{M}_3 \circ
\mathcal{M}_2^{-1} \circ \mathcal{M}_1,
\end{equation}
where:
\begin{subequations}
\begin{eqnarray}
\mathcal{M}_1 \, s & = & s + \frac{1}{2} \Delta \sigma, \\
\mathcal{M}_1 \, p_x & = & p_x + \frac{1}{2} \Delta \sigma \frac{\partial a_s}{\partial x}, \\
\mathcal{M}_2 \, p_x & = & p_x - a_x, \\
\mathcal{M}_2^{-1} \, p_x & = & p_x + a_x, \\
\mathcal{M}_3 \, x & = & x + \Delta \sigma \, p_x.
\end{eqnarray}
\end{subequations}
The effect of the transformations $\mathcal{M}_1$, $\mathcal{M}_2$, $\mathcal{M}_3$,
on variables not explicitly shown is to leave them unchanged.  These transformations provide
a second-order integrator (the error is $O(\Delta \sigma^3)$).

If the components of the vector potential $\vec{a}$ are expressed as power series in
$x$ and $s$, then at this stage we can use a differential algebra code to compose the
transformations first for each step through the field, and then for successive steps through the
field.  However, except for trivial cases, this will require truncation of the power series in order
maintain a manageable number of terms in the power series, and the truncation will mean a
loss of symplecticity.  Furthermore, we know from the symplectic condition that many of the
coefficients in the power series will be related to each other: this direct approach is therefore
more computationally expensive than necessary.

Instead, we can write down mixed-variable generating functions corresponding to each of the individual
transformations.  Using generating functions of the second kind, it can be seen that generators for
the above transformations may be written:
\begin{subequations}
\begin{eqnarray}
F_1(x_1, p_{x2}, s_1, p_{s2}) & = & x_1 p_{x2} - \frac{1}{2} \Delta \sigma a_s(x_1, s_1) + s_1 p_{s2} + \frac{1}{2} \Delta \sigma p_{s2}, \\
F_2(x_1, p_{x2}, s_1, p_{s2}) & = & x_1 p_{x2} + \int_0^{x_1} a_x(x,s_1) \, dx + s_1 p_{s2}, \\
F_3(x_1, p_{x2}, s_1, p_{s2}) & = & x_1 p_{x2} + \frac{1}{2} \Delta \sigma \, p_{x2}^2 + s_1 p_{s2},
\end{eqnarray}
\end{subequations}
where $F_i$ is the generator for the map $\mathcal{M}_i$.  The generator for $\mathcal{M}^{-1}_2$ is:
\begin{equation}
F^{-1}_2(x_1, p_{x2}, s_1, p_{s2}) = x_1 p_{x2} - \int_0^{x_1} a_x(x,s_1) \, dx + s_1 p_{s2}.
\end{equation}
To proceed, we need a way to compose the generating functions.  We discuss two possible methods
for this in Sections \ref{mvgfcomposition1} and \ref{mvgfcomposition2}.
However, we first discuss an integrator, expressed in terms of mixed variable generating
functions, that can be used even in cases where the paraxial approximation is not valid.

%--------------------------------------------------------------------------------------------------------

\subsection{Symplectic Runge-Kutta integrator\label{sec:rungekutta}}

In the previous section, we wrote down expressions for mixed-variable generating functions
for integration steps where the dynamics could be described by a Hamiltonian in the paraxial
approximation.  While the expressions obtained can be expressed rather simply, there are
cases of interest where the paraxial approximation is not valid.  For such cases, at the
cost of rather more complicated formulae, we can use an alternative integration method,
such as a symplectic Runge-Kutta algorithm.

For a Hamiltonian system, a symplectic Runge-Kutta method may be written \cite{hairer}:
\begin{subequations}
\label{symplecticrungekutta1}
\begin{eqnarray}
x_2 & = & x_1 + \Delta \sigma \sum_{i=1}^n b_i \frac{\partial H(x^i, p_x^i)}{\partial p_x}, \\
p_{x2} & = & p_{x1} - \Delta \sigma \sum_{i=1}^n b_i \frac{\partial H(x^i, p_x^i)}{\partial x},
\label{symplecticrungekutta1b}
\end{eqnarray}
\end{subequations}
where the ``intermediate'' variables $(x^i, p_x^i)$ are given by:
\begin{subequations}
\label{symplecticrungekutta2}
\begin{eqnarray}
x^i & = & x_1 + \Delta \sigma \sum_{i=1}^n a_{ij} \frac{\partial H(x^j, p_x^j)}{\partial p_x}, \\
p_x^i & = & p_{x1} - \Delta \sigma \sum_{j=1}^n a_{ij} \frac{\partial H(x^j, p_x^j)}{\partial x}.
\end{eqnarray}
\end{subequations}
The transformations (\ref{symplecticrungekutta1}) and (\ref{symplecticrungekutta2}) are symplectic if:
\begin{equation}
b_i a_{ij} + b_j a_{ji} = b_i b_j,
\end{equation}
for all $i$ and $j$.  A simple case is given by $n = 1$, $a_{11} = 1/2$ and $b_1 = 1$: this
provides a second-order symplectic integrator.  A fourth-order integrator is given \cite{yoshida}
by $n=2$, and:
\begin{subequations}
\label{rungekuttaorder4coeffs}
\begin{eqnarray}
(a_{ij}) & = & \left( \begin{array}{cc}
\frac{1}{4} & \frac{1}{4} - \frac{\sqrt{3}}{6} \\
\frac{1}{4} + \frac{\sqrt{3}}{6} & \frac{1}{4}
\end{array} \right), \\
(b_i) & = & \left( \begin{array}{cc}
\frac{1}{2} & \frac{1}{2} 
\end{array} \right).
\end{eqnarray}
\end{subequations}
The transformations in Eq.\,(\ref{symplecticrungekutta1}) can be derived from a mixed-variable
generating function of the second kind, given by \cite{hairer}:
\begin{equation}
F = x_1 p_{x2} + \Delta \sigma \sum_{i=1}^n b_i H(x^i, p_x^i) -
\Delta \sigma^2 \sum_{i,j=1}^n b_i a_{ij} \frac{\partial H(x^j, p_x^j)}{\partial x} \frac{\partial H(x^j, p_x^j)}{\partial p_x}.
\label{rkmvgf}
\end{equation}
To write the generating function explicitly, it is necessary first to solve for the intermediate
variables $(x^i, p_x^i)$ using Eq.\,(\ref{symplecticrungekutta2}), then use Eq.\,(\ref{symplecticrungekutta1b})
to eliminate $p_{x1}$ (thus expressing the generating function $F$ purely in terms of $x_1$ and $p_{x2}$).

Although each integration step requires the solution of a number of equations in several variables, it
turns out that the solution can be accomplished in a straightforward fashion (using an iterative technique)
with an algebraic code (such as Mathematica), or a DA code.  We have implemented the fourth-order
Runge-Kutta integrator (with coefficients (\ref{rungekuttaorder4coeffs})), using the mixed-variable
generating function (\ref{rkmvgf}) in Mathematica; we
return to the question of solving the systems of equations necessary to construct the mixed-variable
generating function (in particular, Eqs.\,(\ref{symplecticrungekutta2})) in Section \ref{mvgfcomposition2}.

Note that Wu, Forest, and Robin \cite{wuforestrobin} discuss the extension of their method for the
paraxial case, to cases where the paraxial approximation is not valid.  However, this requires (local)
time integration, which adds complexity to the problem.  The advantage of the Runge-Kutta integrator
is that one can adopt a unified approach, using (effectively) path length as the independent variable,
for Hamiltonians of sufficiently general form for most particle tracking applications in accelerators.

%--------------------------------------------------------------------------------------------------------

\subsection{Generating function composition: induction method\label{mvgfcomposition1}}

Let $F_A(x_1,p_{x2})$ and $F_B(x_2,p_{x3})$ be generating
  functions of the second representing the transformations from
  $(x_1,p_{x1})$ to $(x_2,p_{x2})$ and from $(x_2,p_{x2})$ to
  $(x_3,p_{x3})$ respectively.  Our goal is to find a single mixed
  variable generating function $F_C(x_1,p_{x3})$ that represents the
  effect of the combined transformation of $F_A$ and $F_B$.  We first
  note that $F_A(x_1,p_{x2})$ and $F_B(x_2,p_{x3})$ satisfy \cite{goldstein}:
\begin{equation}
\frac{dF_A}{d\sigma} = p_{x1} \dot{x}_1 + x_2 \dot{p}_{x2},
\end{equation}
and
\begin{equation}
\frac{dF_B}{d\sigma} = {p}_{x2} \dot{{x}}_2 + x_3 \dot{p}_{x3}.
\end{equation}
where the dot now signifies differentiation with respect to the independent variable, $\sigma$.
If we define:
\begin{equation}
F_C = F_A - x_2 p_{x2} + F_B, \label{genfunctionsummation}
\end{equation}
we see that $F_C$ satisfies:
\begin{equation}
\frac{dF_C}{d\sigma} = p_{x1} \dot{x}_1  + x_3 \dot{p}_{x3}.
\end{equation}
Therefore, if we express $F_C$ as a function of $x_1$ and $p_{x3}$, then $F_C$ will provide
a mixed variable generating function of the second kind, for a transformation from $(x_1,p_{x1})$
to $(x_3,p_{x3})$.  In general, Eq.\,(\ref{genfunctionsummation}) gives $F_C$ as a function
of $x_1$, $p_{x2}$, ${x}_2$, and $p_{x3}$.  However we can use the equations:
\begin{equation}
x_2 = \frac{\partial F_A}{\partial p_{x2}}, \label{eq:simultaneouspolynomial1}
\end{equation}
and
\begin{equation}
p_{x2} = \frac{\partial F_B}{\partial {x}_2}, \label{eq:simultaneouspolynomial2}
\end{equation}
to eliminate $p_{x2}$ and ${x}_2$.  We are then left with $F_C$ as a function of
$x_1$ and $p_{x3}$ only, as required for a mixed-variable generating function of the second kind.
If $F_A$ and $F_B$ are written as power series in their respective variables, then elimination
of $p_{x2}$ and ${x}_2$ may be performed as follows.
Equations (\ref{eq:simultaneouspolynomial1}) and (\ref{eq:simultaneouspolynomial2}) may be
written in the form:
\begin{equation}
\left\{
\begin{array}{l}
\mathcal{P}_1(x_2, p_{x2}) = 0, \\
\mathcal{P}_2(x_2, p_{x2}) = 0,
\end{array}
\right.
\label{eq:simultaneouspolynomial3}
\end{equation}
where $\mathcal{P}_1$ and $\mathcal{P}_2$ are polynomials of order $N$ in the given arguments.
In general it is difficult to find exact solutions to (\ref{eq:simultaneouspolynomial3}).
However, we only require perturbative solutions to some order in a perturbation parameter
$\varepsilon$, where the dynamical variables are of order $\varepsilon$. Specifically, $\varepsilon=0$
implies $(x_i,p_{xi})=(0,0)$ and we are on the ideal orbit. Thus we propose solving the perturbation
equation:
\begin{equation}
\left\{
\begin{array}{l}
\mathcal{P}_1(x_2, p_{x2}) = O(\varepsilon^{N+1}), \\
\mathcal{P}_2(x_2, p_{x2}) = O(\varepsilon^{N+1}),
\end{array}
\right.
\label{eq:simultaneouspolynomial3p}
\end{equation}

We define the vector $\mathbf{X}$:
\begin{equation}
\mathbf{X} = \left(
\begin{array}{c}
x_2 \\ p_{x2}
\end{array}
\right) .
\label{eq:bigxvector}
\end{equation}
Then we can re-write Eq.\,(\ref{eq:simultaneouspolynomial3p}) as:
\begin{equation}
\sum_{r=0}^{N} \hat{\mathbf{T}}_r ( \overbrace{\mathbf{X},\ldots,\mathbf{X}}^r )
=
O(\varepsilon^{N+1}),
\label{eq:simultaneouspolynomial4}
\end{equation}
where each term in the summation is a vector, each component of which is a monomial of order $r$
in the components of its vector arguments.  Each of the $\hat{\mathbf{T}}_r$ can be regarded as a tensor
of rank $r+1$, so that $\hat{\mathbf{T}}_0$ is a 2-component vector, $\hat{\mathbf{T}}_1$ is a 2$\times$2 matrix,
and so on.  Using an index notation, we would write, for example:
\begin{equation}
\left[ \hat{\mathbf{T}}_3 (\mathbf{X}_1, \mathbf{X}_2, \mathbf{X}_3) \right]_i = 
\sum_{j,k,l=1}^2 \left[ \hat{\mathbf{T}}_3 \right]_{ijkl} \,
\left[ \mathbf{X}_1 \right]_j \,
\left[ \mathbf{X}_2 \right]_k \,
\left[ \mathbf{X}_3 \right]_l.
\end{equation}
The $\hat{\mathbf{T}}_r$ are obtained from $\mathcal{P}_1$ and $\mathcal{P}_2$, and are symmetric
(i.e. $\left[ \hat{\mathbf{T}}_r (\mathbf{X}_1, \dots, \mathbf{X}_r) \right]_i $ is unchanged if any two
of the arguments are interchanged).

We make two assumptions: first, that $\lim_{\varepsilon\to
  0}(\hat{\mathbf{T}}_1)$ is invertible; and second, that
$\hat{\mathbf{T}}_0=O(\varepsilon)$.  Using the first assumption, we can
define:
\begin{equation}
{\mathbf{T}}_k = - \hat{\mathbf{T}}_1^{-1} \hat{\mathbf{T}}_k.
\end{equation}
In Appendix \ref{sec:proof}, it is shown that a solution to Eq.\,(\ref{eq:simultaneouspolynomial4}) is:
\begin{equation}
\mathbf{X} = \sum_{r=1}^{N} \mathbf{c}_r, \label{eq:psolution1}
\end{equation}
where  $\mathbf{c}_1 = {\mathbf{T}}_0$, and for $r \ge 2$:
\begin{equation}
\mathbf{c}_r = \sum_{k = 2}^r \quad\sum_{\substack{m_1,m_2,\ldots,m_k=1 \\ m_1+\cdots+m_k = r}}
{\mathbf{T}}_k(\mathbf{c}_{m_1}, \dots , \mathbf{c}_{m_k}).
\label{eq:psolution2}
\end{equation}
Here the inner summation is performed over all combinations $(m_1,m_2,\ldots,m_k)$
where each $m_i\ge 1$ and $m_1 + \cdots + m_k=r$. Thus (\ref{eq:psolution2}) can be
rewritten as a multiple sum:
\begin{equation}
\mathbf{c}_r = \sum_{k = 2}^r 
\sum_{m_1=1}^{r} \sum_{m_2=1}^{r-m_1} \sum_{m_3=1}^{r-m_1-m_2} 
\ldots \sum_{m_{k-1}=1}^{r-m_1-\cdots-m_{k-2}}
{\mathbf{T}}_k\big(\mathbf{c}_{m_1}, \dots , \mathbf{c}_{m_{k-1}},\mathbf{c}_{r-m_1-\cdots-m_{k-1}}\big).
\label{eq:psolution2alt}
\end{equation}
The $\mathbf{c}_r$ are given inductively, so that:
\begin{eqnarray}
\mathbf{c}_2 & = &  {\mathbf{T}}_2({\mathbf{T}}_0, {\mathbf{T}}_0), \\
\mathbf{c}_3 & = &  {\mathbf{T}}_3({\mathbf{T}}_0, {\mathbf{T}}_0, {\mathbf{T}}_0) + 2 {\mathbf{T}}_2({\mathbf{T}}_0,{\mathbf{T}}_2({\mathbf{T}}_0, {\mathbf{T}}_0)),
\end{eqnarray}
and so on. We note that $\mathbf{c}_r=O(\varepsilon^{r})$.

The above formulae may be generalised to any number of degrees of freedom.  Thus,
this procedure leads, in principle, to explicit expressions for the coefficients in $F_C$
(expressed as a power series) in terms of the coefficients in $F_A$ and $F_B$.  

However, applied to the composition of generating functions, the solution expressed in
equations (\ref{eq:psolution1}) and (\ref{eq:psolution2}) is not purely numerical, since
the coefficients in the polynomial equations (\ref{eq:simultaneouspolynomial3}) must be
expressed in terms of dynamical variables (in the case of one degree of freedom, $x_1$
and $p_{x3}$).  Unfortunately, for all but the simplest cases (low-order maps, in one
degree of freedom), the expressions obtained are too complicated to be of real practical
use.  Even in one degree of freedom, the expressions up to fourth order %\cite{explicitformulae}
are already very complicated.  We therefore consider an iterative procedure for
composing the generating functions, which simplifies the expressions required.

%--------------------------------------------------------------------------------------------------------

\subsection{Generating function composition: iteration method\label{mvgfcomposition2}}

As an alternative to the induction method, we can attempt to solve the equations
involved in generating function composition using iteration.  The formulae required may
be simpler to implement in practice.  Suppose that we
have two mixed variable generating functions, $F_A(x_1,p_{x2})$
and $F_B(x_2,p_{x3})$, which are known.  $F_A$ gives the transformation from $\sigma_1$ to $\sigma_2$:
\begin{eqnarray}
x_2 & = & \frac{\partial F_A}{\partial p_{x2}}, \label{fax2} \\
p_{x1} & = & \frac{\partial F_A}{\partial x_1}, \label{fapx1}
\end{eqnarray}
and $F_B$ gives the transformation from $\sigma_2$ to $\sigma_3$:
\begin{eqnarray}
x_3 & = & \frac{\partial F_B}{\partial p_{x3}}, \label{fbx3} \\
p_{x2} & = & \frac{\partial F_B}{\partial x_2}. \label{fbpx2}
\end{eqnarray}
We wish to find a generating function $F_C(x_1,p_{x3})$ which gives the transformation from $\sigma_1$ to $\sigma_3$:
\begin{eqnarray}
x_3 & = & \frac{\partial F_C}{\partial p_{x3}}, \label{fcx3} \\
p_{x1} & = & \frac{\partial F_C}{\partial x_1}. \label{fcpx1}
\end{eqnarray}
We have seen in Section \ref{mvgfcomposition1} that if the generating functions are expressed
as polynomials in the appropriate variables, then the problem can be reduced to one
of finding solutions to a set of polynomial equations of the form (\ref{eq:simultaneouspolynomial3}).
We can attempt to find a solution iteratively, using Newton's method, as follows.  First, we construct
the matrix $J$:
\begin{equation}
J = \left( \begin{array}{cc}
\frac{\partial \mathcal{P}_1}{\partial x_2} & \frac{\partial \mathcal{P}_1}{\partial p_{x2}} \\
\frac{\partial \mathcal{P}_2}{\partial x_2} & \frac{\partial \mathcal{P}_2}{\partial p_{x2}}
\end{array} \right).
\end{equation}
Then, if we make an initial estimate of the solution, for example:
\begin{equation}
\mathbf{X}_{(1)} =
\left( \begin{array}{c} 0 \\ 0 \end{array} \right),
\end{equation}
(where $\mathbf{X}$ is given by Eq.\,(\ref{eq:bigxvector})) then we can construct improved estimates:
\begin{equation}
\mathbf{X}_{(n+1)} =
\mathbf{X}_{(n)} -
J_{(n)}^{-1} \left( \begin{array}{c} \mathcal{P}_1 \\ \mathcal{P}_2 \end{array} \right)_{\!\!\!\!(n)}
\end{equation}
starting from $n = 1$.  For Eq.\,(\ref{eq:simultaneouspolynomial3}), the solution needs to be
performed using an algebraic code, since the polynomials are expressed in terms of more
variables than just those for which a solution is sought.  However, we find in practice
(e.g. for the illustration presented in Section \ref{example}), that the iteration can be
carried out quickly and converges rapidly, even when working to high order in the variables.

This approach to solving a system of nonlinear equations can also be applied to the
Runge-Kutta integrator described in Section \ref{sec:rungekutta}.  If the Hamiltonian is
expressed as a polynomial in the dynamical variables (note that this means expanding
the square root in Eq.\,(\ref{hamiltonian1dof}), but this can be done to higher order than
is done for the paraxial approximation), then Eq.\,(\ref{symplecticrungekutta2}) represents $2n$
polynomial equations in the $2n$ ``intermediate''  variables $(x^i, p_x^i)$.  These may be solved iteratively
using Newton's method (in an algebraic code), to express the intermediate variables as power series
in the initial variables $(x_1, p_{x1})$.  Then, it is only necessary to make a substitution of $p_{x2}$
for $p_{x1}$, using Eq.\,(\ref{symplecticrungekutta1b}), to produce an expression for the mixed-variable
generating function (\ref{rkmvgf}) for a single integration step.

%--------------------------------------------------------------------------------------------------------

\section{Example: quadrupole with octupole component and fringe field\label{example}}

\begin{figure}
\begin{center}
\includegraphics[width=0.6\textwidth]{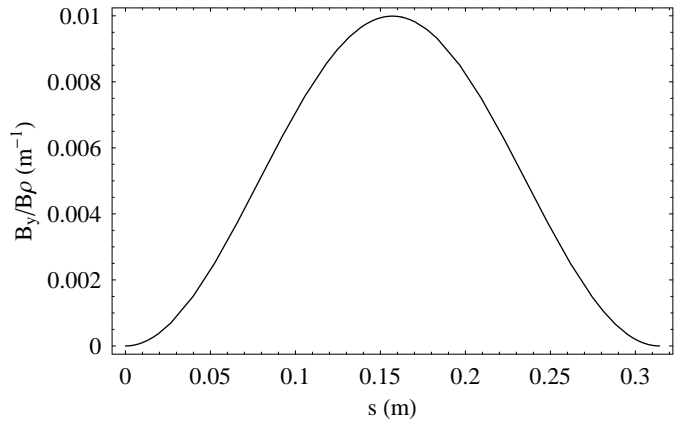}
\end{center}
\caption{Vertical normalised field component in the example magnet, along the line $x = -1$\,mm,
$y = 0$ from $s = 0$ to $s = L$.\label{normalisedfield}}
\end{figure}

As an example, we illustrate the construction of the map, in the form of a mixed-variable
generating function of the second kind, for a quadrupole with strong octupole component.  The
axis of the magnet (on which the field is zero) defines the reference trajectory.  We allow both the
quadrupole and octupole components to vary with position along the axis of the magnet, so as to
represent some fringe field region at the entrance and exit of the magnet. An appropriate
representation of the field, including the longitudinal variation, is given by Dragt's generalised
gradients \cite{dragtgeneralisedgradients}.  The appropriate expressions are given in
reference \cite{dragtgeneralisedgradients2}; for convenient reference, they are reproduced in
Appendix \ref{sec:generalisedgradients}.  In the notation used there, our field is defined by:
\begin{eqnarray}
C_2(s) & = & \frac{1}{2} \frac{k_1}{B\rho} \sin^2(k_s s), \label{generalisedgradient1} \\
C_4(s) & = & \frac{1}{4!} \frac{k_3}{B\rho} \sin^2(k_s s), \label{generalisedgradient3} 
\end{eqnarray}
with $k_1 = -10$\,m$^{-2}$, $k_3 = 6\times 10^4$\,m$^{-4}$, and $k_s = 10$\,m$^{-1}$.
The total length of the magnet is defined to be $L = \pi / k_s$.
The expressions for the vector potential involve infinite summations: in practice, these must be
truncated at some order in the co-ordinates.  Although the fields corresonding to the truncated
potentials strictly speaking do not satisfy Maxwell's equations, the error is of order higher than the
truncation; and if the truncation is at significantly high order, the transfer map, itself computed
only up to some order, is not affected.  Note that the expressions for the transformations in the
symplectic integrator involve the vector potential normalised by $P_0/q$: therefore, it is not
necessary to specify either the particle charge or the reference momentum.
The normalised vertical field along the line $x = -1$\,mm, $y = 0$ from $s = 0$ to $s = L$ is
shown in figure \ref{normalisedfield}.

Using the generalised gradients (\ref{generalisedgradient1}) and (\ref{generalisedgradient3}), we
compute the mixed-variable generating function for the transfer map from $s=0$ to $s=L$, using
the fourth-order Runge-Kutta integrator (Section \ref{sec:rungekutta}), and the iterative method for
composing the mixed-variable generating functions (Section \ref{mvgfcomposition2}).  The Hamiltonian
is given by Eq.\,(\ref{hamiltonian1dof}), with the square root expanded to sixth order in the dynamical variables.
Terms up to sixth order in the co-ordinates are retained in the expression for the vector potential.
We retain terms to fourteenth order in the dynamical variables when composing the generating function,
to reduce feed-down errors.
A step size of $\Delta\sigma = L/1024$ was used in the integration.
All computations were performed in Mathematica.

\begin{figure}
\begin{center}
\includegraphics[width=0.48\textwidth]{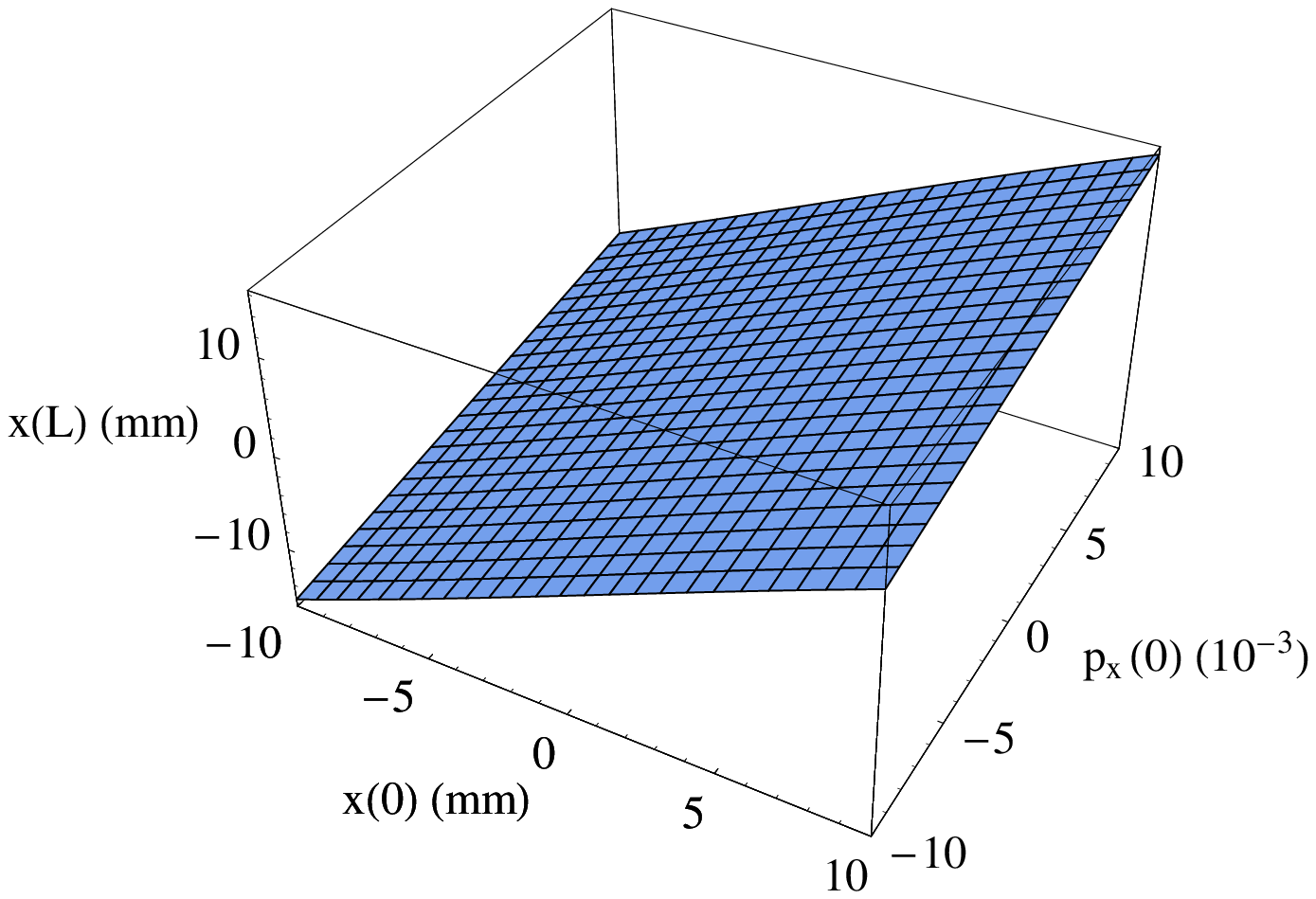}
\includegraphics[width=0.48\textwidth]{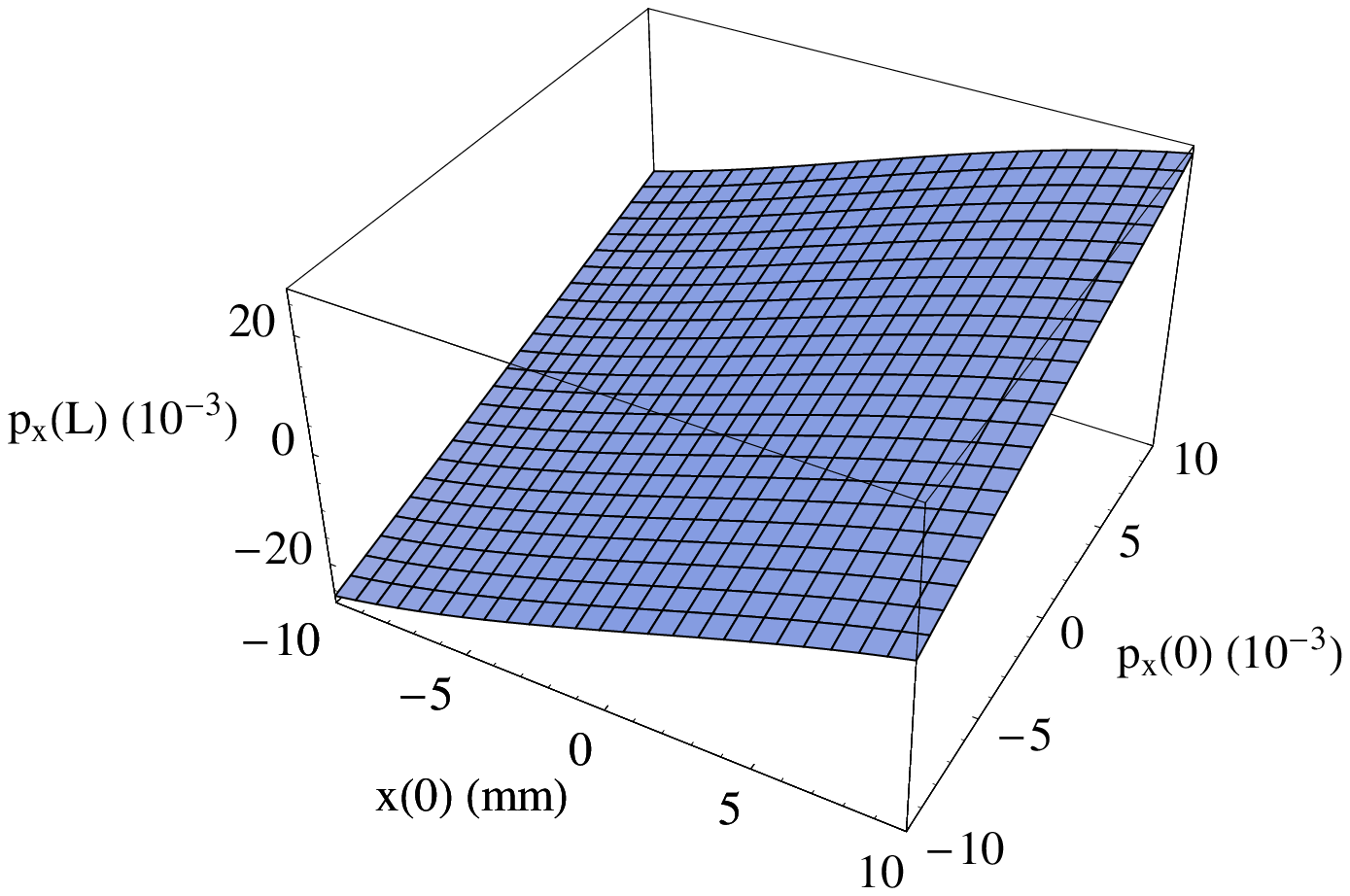}
\includegraphics[width=0.48\textwidth]{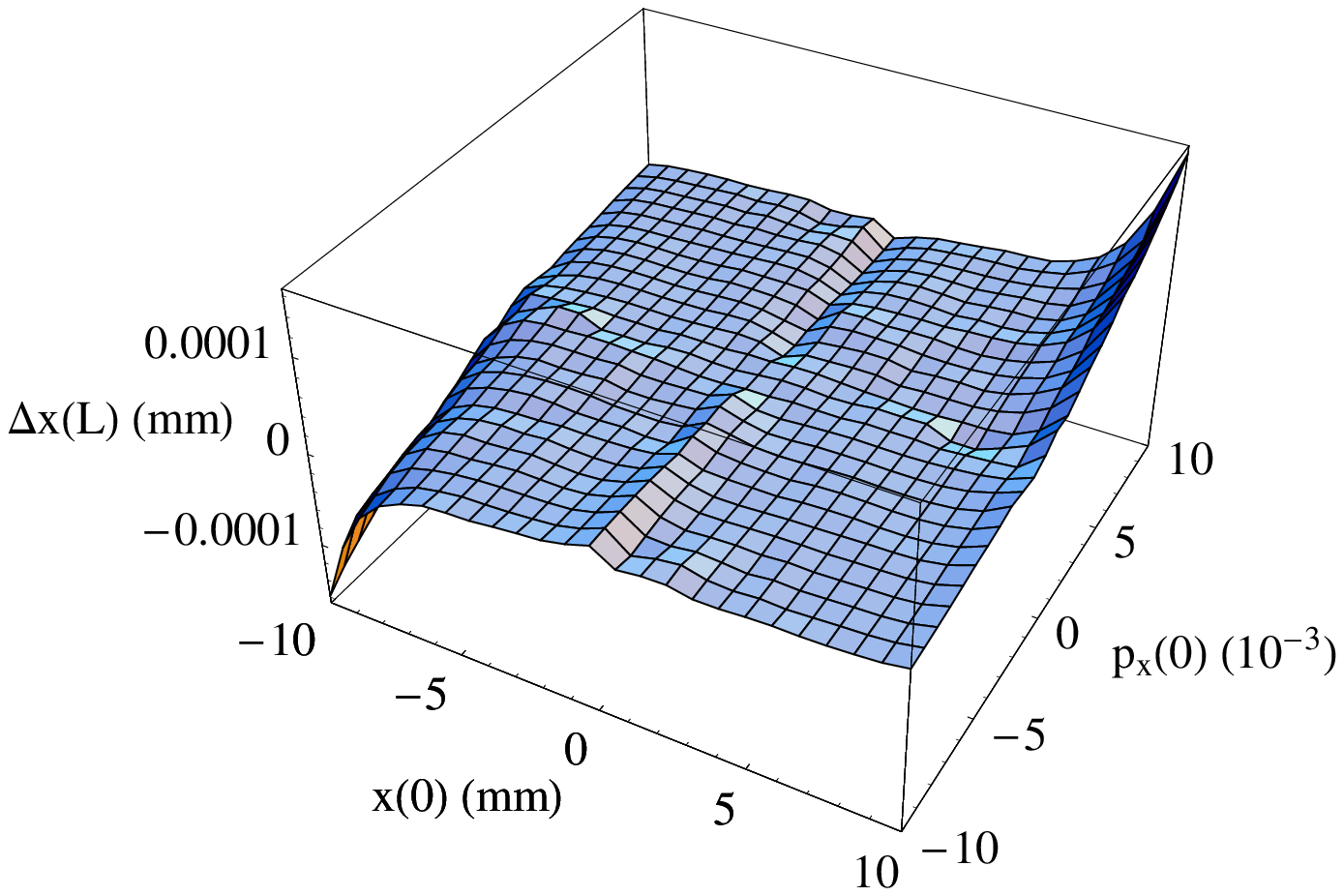}
\includegraphics[width=0.48\textwidth]{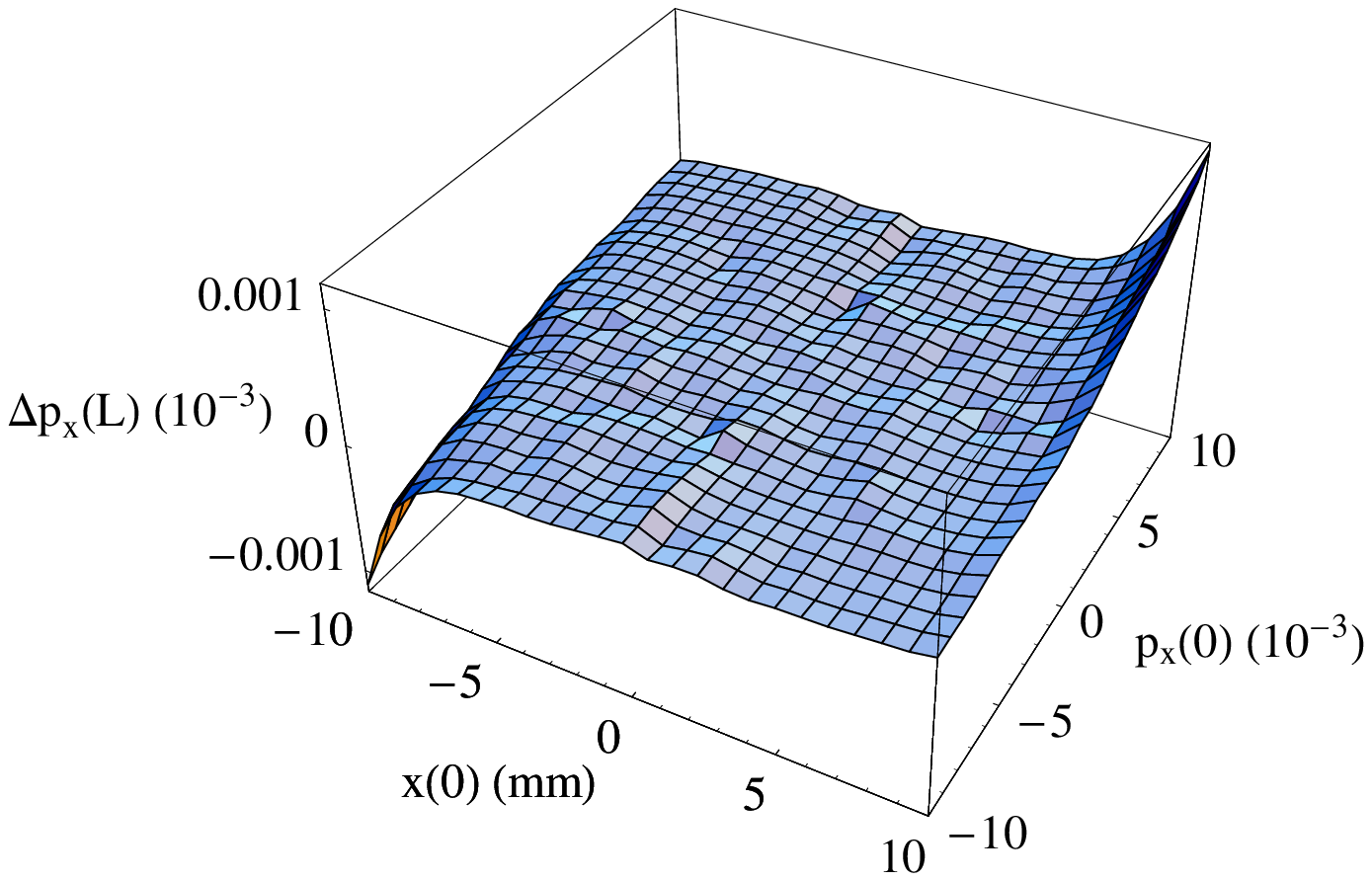}
\end{center}
\caption{Transfer functions in the example magnet.  Top: horizontal co-ordinate (left) and normalised
momentum (right) at $s = L$ as functions of co-ordinate and momentum at $s=0$.
Bottom: residuals for the final horizontal co-ordinate (left) and normalised momentum (right), between
the values calculated using the mixed-variable generating function, and a purely numerical integration
through the field.\label{transferfunctions}}
\end{figure}

The transfer functions (final co-ordinate and momentum, as functions of initial co-ordinate and momentum)
are shown in figure \ref{transferfunctions}.  For comparison, we also computed the transfer functions
using direct numerical integration of Hamilton's equations, using the Hamiltonian (\ref{hamiltonian1dof})
(without any approximation for the square root).
The residuals between the transfer functions obtained from the mixed-variable generating function,
and the transfer functions obtained from the numerical integration are also shown in
figure \ref{transferfunctions}.
 
\begin{figure}
\begin{center}
\includegraphics[width=0.48\textwidth]{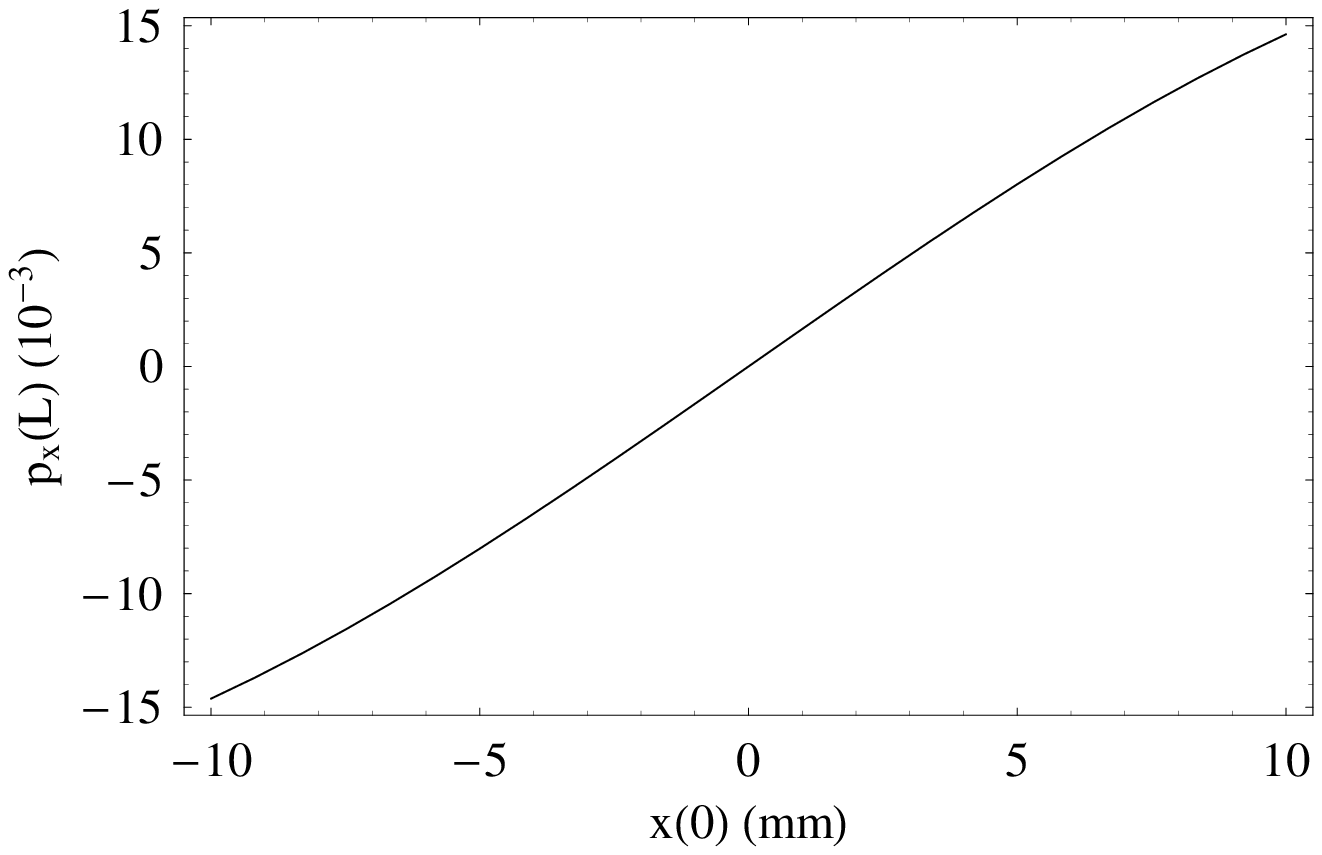}
\includegraphics[width=0.48\textwidth]{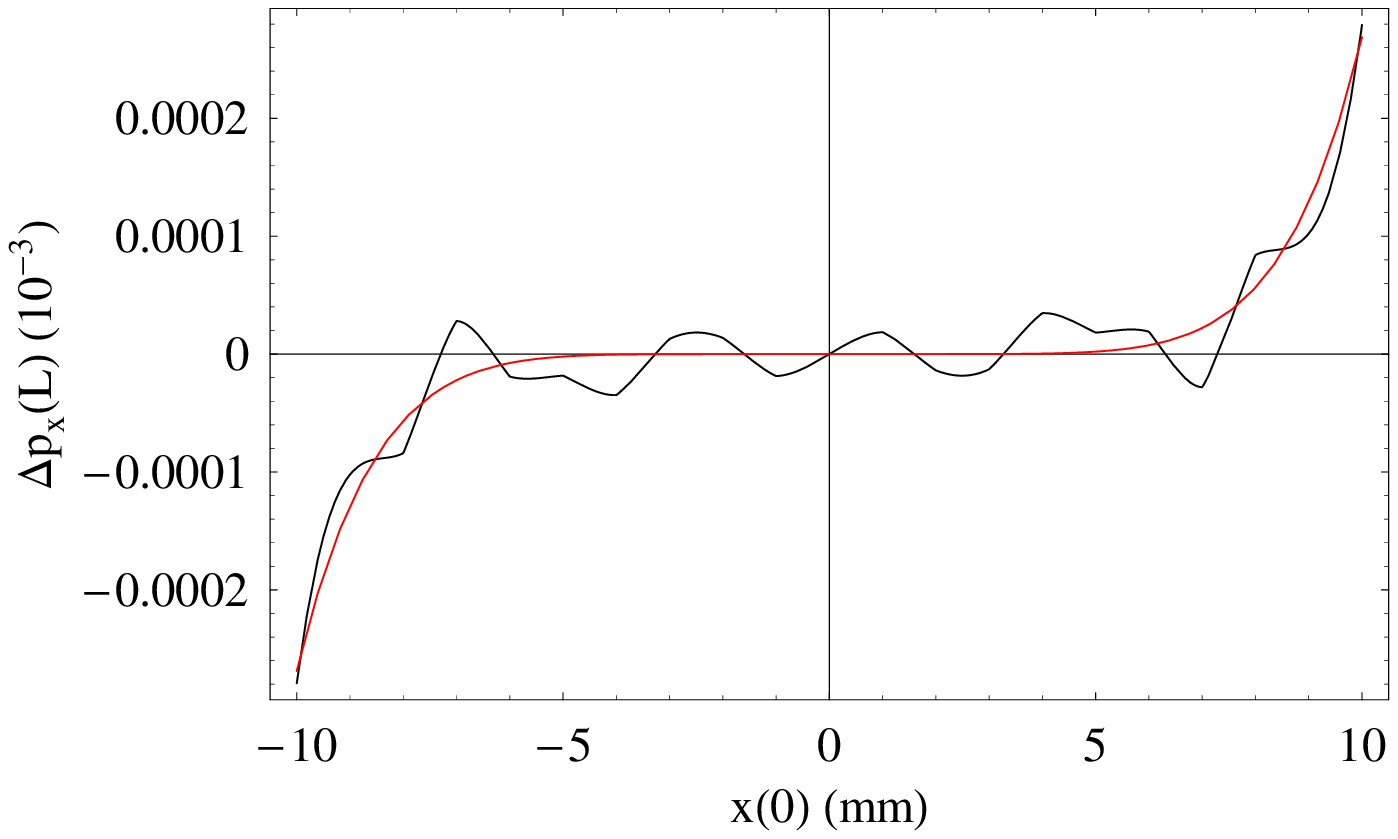}
\caption{Left: normalised horizontal momentum at $s = L$, as a function of horizontal co-ordinate at
$s = 0$, for $p_x(0)=0$.  Right: residual in final normalised horizontal momentum between values
calculated from the mixed-variable generating function and from direct numerical integration.
The black line shows the residual, the red line shows a fit to the residual, using a function of the
form $\Delta p_x(L) \propto x(0)^7$. \label{transferfunction1d}}
\end{center}
\end{figure}

The residuals appear small over the range of variables plotted.  To investigate the residuals more
closely, we look specifically at the value of $p_x(L)$ as a function of $x(0)$, for $p_x(0) = 0$: we refer
to this as the 1-D (one-dimensional) transfer function.  A plot of the 1-D transfer function is shown in
figure \ref{transferfunction1d} (note that there is some third-order curvature evident, from the octupole
component in the quadrupole).   Also shown in figure \ref{transferfunction1d} is the difference (the
residual) between the 1-D transfer functions obtained from the mixed-variable generating function and
from direct numerical integration.  The residual can be fitted (apart from some fluctuations, which
may be due to limits on numerical precision) using a polynomical with only a single term, in $x^7$.
This indicates that the residual is dominated by the seventh-order terms in the map.

We can also compare the coefficients in the 1-D transfer function (shown in figure \ref{transferfunction1d},
left):
\begin{equation}
p_x(L) = \sum_{m=1}^5 h_m x(0)^m.
\end{equation}
The coefficients $h_m$ can be obtained using various techniques; a comparison of the results from some
methods are shown in table \ref{table1}.  First, we obtain the $h_m$ directly from the generating
function, by solving:
\begin{equation}
p_{x1} = 0 = \frac{\partial F}{\partial x_1}, \label{mvgf1dxferfn}
\end{equation}
where $F$ is the generating function for a map through the entire magnet; the solution may be expanded
in a power series in $x_1$, which gives the $h_m$ directly.  Second, we can obtain the $h_m$ from
a polynomial fit to the 1-D transfer function obtained by numerical integration.  Finally, estimated values
for the $h_m$ may be obtained by integrating the appropriate normalised multipole component along the
reference trajectory:
\begin{equation}
h_m \approx -\frac{1}{m!} \int k_m \, ds. \label{fieldintegral}
\end{equation}
The results from the first two methods are in good agreement up to the fourth-order ($h_4$), and in
reasonable agreement for $h_5$.  The values obtained using the third method (integrating the
normalised multipole component along the reference trajectory, Eq.\,(\ref{fieldintegral})) are only
approximate: effects associated with the fringe fields, and the non-zero
length of the magnet, are not properly taken into account.

\begin{table}
\caption{Coefficients in the transfer function, derived from the mixed-variable generating function
(MVGF), and by a polynomial fit to the tracking results using direct
numerical integration (NI). Also shown for comparison
are the values that would be expected from the field integrals (FI).
\label{table1}}
\begin{center}
\begin{tabular}{l c c c}
\hline
m & $h_m$ & $h_m$ & $h_m$ \\
   & MVGF & NI & FI \\ 
\hline
1 &  1.65228\,m$^{-1}$                            & 1.65226\,m$^{-1}$ & 1.5708\,m$^{-1}$ \\
2 & -1.00075$\times 10^{-13}$\,m$^{-2}$ & -1.00075$\times 10^{-13}$\,m$^{-2}$ & 0 \\
3 & -1933.15\,m$^{-3}$                            & -1930.82\,m$^{-3}$ & -1570.8\,m$^{-3}$ \\
4 &  2.21872$\times 10^{-9}$\,m$^{-4}$   &  2.21798$\times 10^{-9}$\,m$^{-4}$ & 0 \\
5 &  3.84174$\times 10^{5}$\,m$^{-5}$    & 3.30479$\times 10^{5}$\,m$^{-5}$ & 0 \\
\hline
\end{tabular}
\end{center}
\end{table}

\begin{figure}
\includegraphics[width=0.48\textwidth]{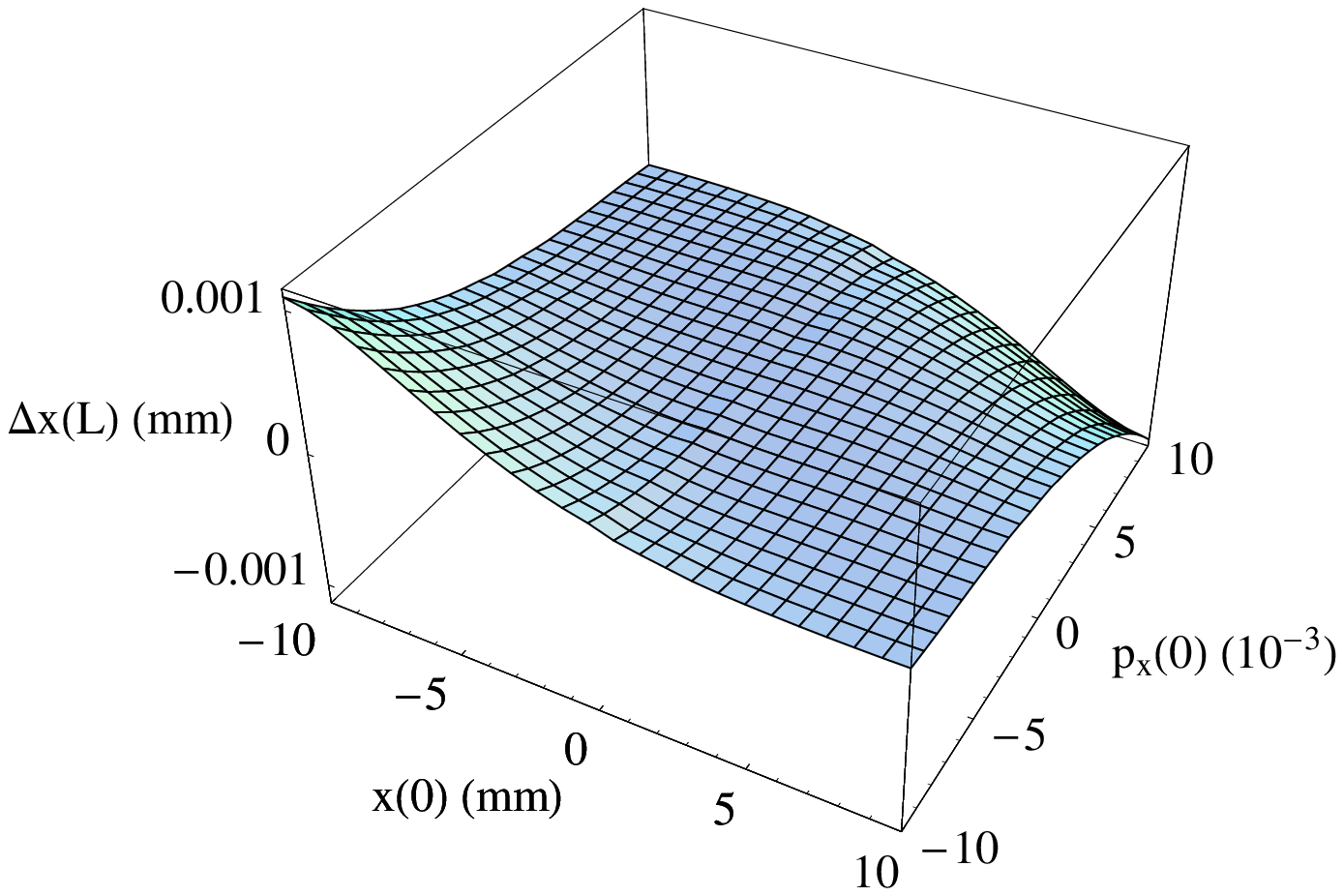}
\includegraphics[width=0.48\textwidth]{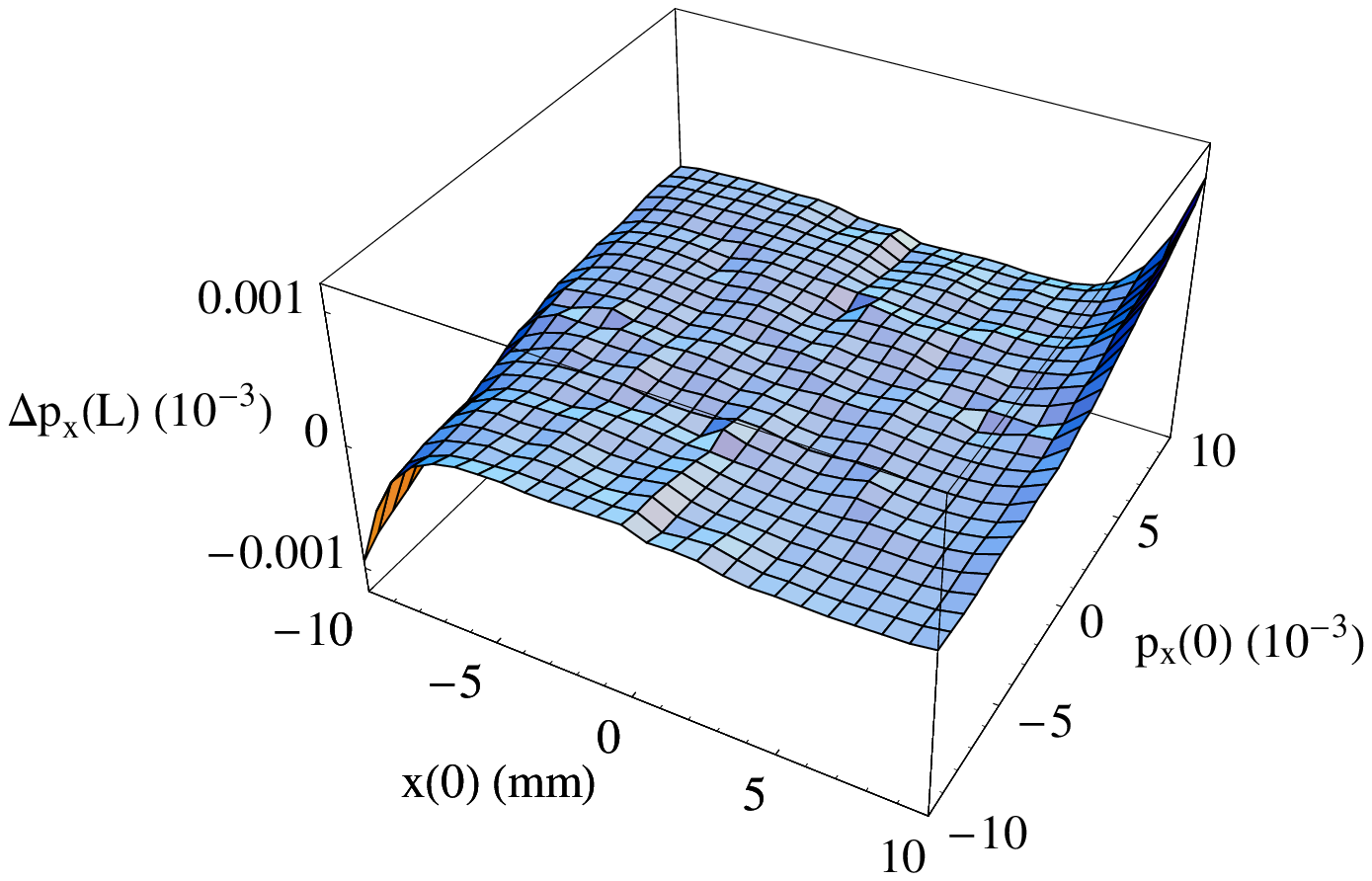}
\caption{Residuals in the final co-ordinate (left) and final momentum (right) as a function of initial co-ordinate
and momentum.  The residuals are calculated as the difference between the results found from the mixed-variable
generating function for the ``exact'' Hamiltonian, and the
numerical integration of the Hamiltonian in the paraxial approximation.
\label{paraxialeffect}}
\end{figure}

Finally, we can see the effect the paraxial approximation would have if applied to this example, by
comparing the residuals between the results of the direct numerical integration using the Hamiltonian
(\ref{hamiltonian1dof}), and the Hamiltonian in the paraxial approximation (\ref{hamiltonianparaxial1dof}).
The difference between the two is small in this case, but visible if we compare the difference in the
final co-ordinate, as a function of initial co-ordinate and momentum: this is shown in figure \ref{paraxialeffect}.
The residuals are an order of magnitude larger than those between the direct numerical integration for the
``exact'' Hamiltonian, and the integration (with mixed-variable generating functions) for the Hamiltonian
expanded to sixth order in the dynamical variables.  This suggests that using mixed-variable generating
functions, we can readily construct an exactly symplectic high-order map to good accuracy, without needing
to make the paraxial approximation in the Hamiltonian for the system.

%--------------------------------------------------------------------------------------------------------

%\section{Final remarks}

%--------------------------------------------------------------------------------------------------------

\appendix

%--------------------------------------------------------------------------------------------------------

\section{Proof of the inductive formula for solving a system of polynomial equations}
\label{sec:proof}

We wish to show that (\ref{eq:psolution2}) is a solution to
(\ref{eq:simultaneouspolynomial4}). 
Assuming that $\hat{\mathbf{T}}_1$ is invertible, we multiply
Eq.\,(\ref{eq:simultaneouspolynomial4}) by $-\hat{\mathbf{T}}_1^{-1}$
to give:
\begin{equation}
\sum_{r=2}^{N} {\mathbf{T}}_r ( \mathbf{X},\ldots,\mathbf{X} ) - \mathbf{X} + {\mathbf{T}}_0 = O(\varepsilon^{N+1}).
\label{eq:a1}
\end{equation}
Let $\mathbf{S}_n$ be the left hand side of (\ref{eq:a1}):
\begin{equation}
\mathbf{S}_n
=
\sum_{r=2}^{N} {\mathbf{T}}_{r}(\overbrace{\mathbf{X},\ldots,\mathbf{X}}^r)
-
\mathbf{X}
+{\mathbf{T}}_0.
\end{equation}
Substituting for $\mathbf{X}$ from Eq.\,(\ref{eq:psolution1}) gives:
\begin{equation}
\mathbf{S}_n
=
\sum_{r=2}^{N} {\mathbf{T}}_{r}\Big(\sum_{m_1=1}^\infty\mathbf{c}_{m_1},
\ldots,\sum_{{m_r}=1}^\infty\mathbf{c}_{m_r}\Big)
-
\sum_{r=1}^\infty\mathbf{c}_r
+{\mathbf{T}}_0.
\end{equation}
Since $\mathbf{c}_1={\mathbf{T}}_0$, we find:
\begin{eqnarray}
\mathbf{S}_n
&=&
\sum_{r=2}^{N} {\mathbf{T}}_{r}\left(\sum_{m_1=1}^{N}\mathbf{c}_{m_1},
\ldots,\sum_{{m_r}=1}^{N}\mathbf{c}_{m_r}\right)
-
\sum_{r=2}^{N}\mathbf{c}_r
+ O(\varepsilon^{N+1})
\\&=&
\sum_{r=2}^{N} \sum_{m_1=1}^{N}\cdots\sum_{{m_r}=1}^{N}
{\mathbf{T}}_{r}
\big(\mathbf{c}_{m_1},
\ldots,\mathbf{c}_{m_r}\big)
-
\sum_{r=2}^{N}\mathbf{c}_r
+ O(\varepsilon^{N+1})
\\&=&
\sum_{r=2}^{N} \sum_{k=2}^{N} 
\sum_{\substack{m_1,\ldots,m_k=1 \\ m_1+\cdots+m_k=k}}
{\mathbf{T}}_{r}
\big(\mathbf{c}_{m_1},
\ldots,\mathbf{c}_{m_r}\big)
-
\sum_{k=2}^{N}\mathbf{c}_k
+ O(\varepsilon^{N+1})
\\&=&
\sum_{k=2}^{N}
\bigg(-\mathbf{c}_k
+
\sum_{r=2}^{N} 
\sum_{\substack{m_1,\ldots,m_k=1 \\ m_1+\cdots+m_k=k}}
{\mathbf{T}}_{r}
\big(\mathbf{c}_{m_1},
\ldots,\mathbf{c}_{m_r}\big)
\bigg)
+ O(\varepsilon^{N+1}).
\end{eqnarray}
For $r>k$ the set $\big\{(m_1,\ldots,m_r)\,\big|\,m_i\ge 1\text{ and }\,m_1+\ldots+m_r=k\big\}=\emptyset$, hence:
\begin{equation}
\mathbf{S}_n
=
\sum_{k=2}^{N}
\bigg(-\mathbf{c}_k
+
\sum_{r=2}^{k} 
\sum_{\substack{m_1,\ldots,m_k=1 \\ m_1+\cdots+m_k=k}}
\big(\mathbf{c}_{I_1},
\ldots,\mathbf{c}_{I_n}\big)
\bigg)
+ O(\varepsilon^{N+1}).
\end{equation}
Substituting for $\mathbf{c}_k$ from Eq.\,(\ref{eq:psolution2}) gives:
\begin{equation}
\mathbf{S}_n=O(\varepsilon^{N+1}),
\end{equation}
which is Eq.\,(\ref{eq:simultaneouspolynomial4}).

%--------------------------------------------------------------------------------------------------------

\section{Generalised gradients\label{sec:generalisedgradients}}

In Cartesian variables, a vector potential in the Coulomb gauge can be expressed as
\cite{dragtgeneralisedgradients2}:
\begin{eqnarray}
A_x & = & \frac{1}{2} \sum_{m=1}^\infty \Re (x + iy)^{m+1}
\sum_{\ell=0}^\infty (-1)^\ell \frac{m!}{2^{2\ell} \ell ! (\ell + m + 1)!}
C^{[2\ell + 1]}_m (z) (x^2 + y^2)^\ell , \\
A_y & = & \frac{1}{2} \sum_{m=1}^\infty \Im (x + iy)^{m+1}
\sum_{\ell=0}^\infty (-1)^\ell \frac{m!}{2^{2\ell} \ell ! (\ell + m + 1)!}
C^{[2\ell + 1]}_m (z) (x^2 + y^2)^\ell , \\
A_z & = & - \sum_{m=1}^\infty \Re (x + iy)^{m}
\sum_{\ell=0}^\infty (-1)^\ell \frac{m!}{2^{2\ell} \ell ! (\ell + m)!}
C^{[2\ell]}_m (z) (x^2 + y^2)^\ell,
\end{eqnarray}
where:
\begin{equation}
C^{[k]}_m = \frac{d^k C_m}{dz^k}.
\end{equation}
Maxwell's equations in free space are satisfied for any differentiable functions $C_m(z)$, which
are called the generalised gradients.
The conventional multipole expansion is obtained in the case that the $C_m(z)$ are independent
of $z$; then, $C_1$ represents the dipole field component, $C_2$ the quadrupole field component,
and so on.

The symplectic integrator described in Section \ref{symplecticintegrator} uses a vector
potential in the gauge:
\begin{equation}
A_y = 0.
\end{equation}
This is readily obtained from the above expressions using the gauge transformation:
\begin{equation}
\vec{A}^\prime = \vec{A} - \nabla \psi,
\end{equation}
where:
\begin{equation}
\psi = \int A_y \, dy.
\end{equation}

%--------------------------------------------------------------------------------------------------------

\begin{acknowledgments}
We should like to thank B. Muratori for suggesting the use of
symplectic Runge-Kutta integration for particle tracking.
This work was supported by the Science and Technology Facilities Council, UK.
\end{acknowledgments}

%--------------------------------------------------------------------------------------------------------

\end{document}